\documentclass[lettersize,journal]{IEEEtran}
\usepackage{blindtext,cite,hyperref}
\usepackage{ctable,amsmath,amssymb,amsfonts}
\usepackage{algorithmic}
\usepackage{algorithm}
\usepackage{graphicx}
\usepackage{textcomp}
\usepackage{makecell}
\usepackage{float}
\usepackage{placeins}
\usepackage{caption}
\usepackage{pifont}
\begin{document}

\title{A Unifying Approach to Inverse Problems of Ultrasound Beamforming and Deconvolution}
\author{Sobhan Goudarzi, Adrian Basarab, \IEEEmembership{Senior Member, IEEE}, Hassan Rivaz, \IEEEmembership{Senior Member, IEEE}
\thanks{This work was supported by Natural Sciences and Engineering Research Council of Canada (NSERC) RGPIN-2020-04612.}
\thanks{Sobhan Goudarzi and Hassan Rivaz are with the Department of Electrical and Computer Engineering, Concordia University, Montreal, QC, H3G 1M8, Canada.\newline
	Adrian Basarab is with the Universit\'e de Lyon, INSA-Lyon, UCBL, CNRS, Inserm, CREATIS UMR 5220, U1206, Villeurbanne, France.\newline
	Email: sobhan.goudarzi@concordia.ca~, adrian.basarab@creatis.insa-lyon.fr~, hrivaz@ece.concordia.ca~}}
\maketitle
%%%%%%%%%%%%%%%%%%%%%%%%%%%%%%%%%%%%%%%%%%%%%%%%%%%%%%%%%%%%%%%%%%%%%%%%%%%%%%%%%%%%%%%%%%%%%%%%%%%%%%%%
\begin{abstract}
Beamforming is an essential step in the ultrasound image formation pipeline and has recently attracted growing interest. An important goal of beamforming is to increase the image spatial resolution, or in other words to narrow down the system point spread function. In parallel to beamforming approaches, deconvolution methods have also been explored in ultrasound imaging to mitigate the adverse effects of PSF. Unfortunately, these two steps have only been considered separately in a sequential approach. Herein, a novel framework for unifying beamforming and deconvolution in ultrasound image reconstruction is introduced. More specifically, the proposed formulation is a regularized inverse problem including two linear models for beamforming and deconvolution plus additional sparsity constraint. We take advantage of the alternating direction method of multipliers algorithm to find the solution of the joint optimization problem. The performance evaluation is presented on a set of publicly available simulations, real phantoms, and \textit{in vivo} data. Furthermore, the superiority of the proposed approach in comparison with the sequential approach as well as each of the other beamforming and deconvolution approaches alone is also shown. Results demonstrate that our approach combines the advantages of both methods and offers ultrasound images with superior resolution and contrast.  
\end{abstract}
%%%%%%%%%%%%%%%%%%%%%%%%%%%%%%%%%%%%%%%%%%%%%%%%%%%%%%%%%%%%%%%%%%%%%%%%%%%%%%%%%%%%%%%%%%%%%%%%%%%%%%%%
%%%%%%%%%%%%%%%%%%%%%%%%%%%%%%%%%%%%%%%%%%%%%%%%%%%%%%%%%%%%%%%%%%%%%%%%%%%%%%%%%%%%%%%%%%%%%%%%%%%%%%%%
\section{Introduction}
\label{sec:sec1}
\IEEEPARstart{S}{onography} is among the most prevalent medical imaging modalities and has the advantages of being cost-effective, non-invasive, portable, and real-time. Ultrasound imaging, however, suffers from several artifacts such as clutter and poor resolution, which emanate from its wide Point Spread Function (PSF). There are several approaches to tackle this problem, among which beamforming~\cite{165556} and deconvolution~\cite{1539261} are the foci of the current work.\par
Beamforming is an essential step in the medical ultrasound image formation pipeline. In the transmission step, beamforming specifies the firing time, as well as the shape of the excitation pulse, applied to each element of the transducer in order to create the desired acoustic wave~\cite{cobbold2006}. In receive step, beamforming is applied to trace back the backscattered echoes from each voxel of the medium~\cite{cobbold2006}. Receive beamforming can be accomplished in time~\cite{4816058} and frequency~\cite{655200,1539703,6863846} domains. Delay-and-sum (DAS) beamforming is the most common time-domain approach due to its simplicity and efficacy. To reconstruct the spatial map of the target echogenicity using DAS, first, the received signals of transducer elements are aligned by applying time delays equal to the two-way propagation times of the transmitted wave reaching a scatterer and getting back to the transducer elements. Afterward, delayed signals are merged using a set of predefined apodization weights. However, there is always a trade-off between the width of the main-lobe and the level of side-lobes in classical apodization windows in the frequency domain. Adaptive methods have been extended to optimize the apodization weights based on the received channel data. They have shown the capability of providing significant improvements in lateral resolution and contrast of B-mode images~\cite{5278437,849225,6960091}.\par
Ultrasound beamforming has also been formulated as an inverse problem in recent years~\cite{7565515,8052532,7728907}. In this formulation, the desired ultrasound image is directly estimated from the observation signals without calculating apodization weights. This framework allows the consideration of any additional constraint, such as Gaussian and Laplacian statistics, in the recovery of desired data.\par
Although a better beamformer improves the system PSF, ultrasound imaging still has a non-ideal PSF due to many factors such as the limited bandwidth of piezoelectric crystal elements, the physical phenomena of acoustic wave propagation in the tissue, etc. Under the assumption of weak scattering for soft tissues and using the first-order Born approximation, the ultrasound Radio-Frequency (RF) data can be linearly modeled as the result of convolution between the ground-truth Tissue Reflectivity Function (TRF) and the PSF of the ultrasound imaging system~\cite{1539261,jensen1993}. Therefore, another line of research has been devoted to mitigating the adverse effect of non-ideal PSF using the deconvolution approach~\cite{1539261,4376242,7302609}. To our knowledge, deconvolution has only been applied as a post-processing approach after reconstructing the ultrasound image based on existing beamforming procedures, essentially DAS.\par
The current study is motivated by the novel idea of unifying the beamforming and deconvolution steps together to simultaneously take advantage of both methods in image reconstruction. The proposed framework is a joint inverse problem including two linear models of beamforming and deconvolution plus an additional sparsity constraint. In other words, our method is designed to estimate the desired image directly and concurrently minimize the adverse effect of the PSF. The resulting optimization problem is solved using split-variable Alternating Direction Method of Multipliers (ADMM) algorithm~\cite{2200000016,6954405} as it allows the minimization of each term of the objective function separately. The performance evaluation is completed on a set of publicly available simulations, real phantoms, and \textit{in vivo} data from the Plane-wave Imaging Challenge in Medical UltraSound (PICMUS)~\cite{7728908}. Furthermore, the superiority of the proposed approach in a comprehensive comparison with the sequential approach and each of the beamforming and deconvolution approaches alone is also shown.
%%%%%%%%%%%%%%%%%%%%%%%%%%%%%%%%%%%%%%%%%%%%%%%%%%%%%%%%%%%%%%%%%%%%%%%%%%%%%%%%%%%%%%%%%%%%%%%%%%%%%%% 
\subsection{Related work}
\label{sec:sec11}
In the rich body of literature on the topic of ultrasound beamforming and deconvolution, the most recent and relevant studies are reviewed in this section.\\*
Beamforming approaches can be categorized into four main groups. The first set is time-domain methods, among which DAS is the most popular non-adaptive method. Filtered-delay multiply and sum (F-DMAS) was proposed in~\cite{6960091}. The algorithm is based on a pairwise multiplication of delayed RF signals before summation. Another extension to DAS has recently been proposed based on null subtraction imaging~\cite{8493541}, wherein envelope images reconstructed by different apodization windows are linearly combined in order to overcome the classical trade-off and have both a low side-lobe level and a narrow main-lobe.\par
Minimum Variance beamforming (MVB)~\cite{4291510} is the most potent approach among adaptive algorithms that mainly improves the lateral resolution perpendicular to the wave propagation called axial direction in ultrasound imaging~\cite{6932199}. In MVB, the main challenge is estimating the covariance matrix from the data, making it computationally expensive~\cite{5278437}. The MVB has also been extended using singular value decomposition (SVD) of the covariance matrix to improve the contrast~\cite{5611687}. But there is no clear criterion for removing small eigenvalues, and a part of speckle texture might be omitted in this method. To speed up MVB, Nilsen and Hafizovic~\cite{5306765} proposed beamspace MVB. Their method is based on the extraction of spatial statistics from a set of orthogonal beams formed in a different direction. This idea has later been used to extend MVB using multibeam covariance matrices~\cite{6217562}. A fast version of MVB has been developed based on principal component analysis (PCA)~\cite{6819209} as well as Legendre polynomials~\cite{7516704}. In~\cite{GOUDARZI2021106036}, the apodization weights are estimated using independent component analysis (ICA).\par
There is another type of adaptive method, in the time domain, based on the coherence factor (CF), which is defined as the ratio of coherent to incoherent energy across the aperture~\cite{mallart1994}. CF was used as an adaptive weight on top of DAS to improve the image quality~\cite{849225}. Generalized CF (GCF) was derived from the spatial spectrum and defined as the ratio between the energy of a predefined low-frequency range (the coherent portion of RF data) to the total spectral energy~\cite{1182117}. Subsequently, phased CF (PCF) was proposed based on the phase, rather than amplitude information of aperture~\cite{4976281}.\par
The second group of beamforming methods is implemented using the Fourier transform. The pioneer studies were based on synthetic aperture focusing~\cite{1539703,139112}. Later on, the Fourier beamforming was extended for plane-wave Imaging~\cite{655200} and implemented through different strategies~\cite{6587395,6932218,6373791,8359331}. Wagner~\textit{et al.} proposed compressed beamforming that works on the sub-Nyquist RF data~\cite{6203608}. Consequently, this idea has been extended as a general beamformer in Fourier domain~\cite{6863846,7582552,8411460}.\par
Szasz~\textit{et al.} proposed the third group of beamforming methods~\cite{7565515}, which assume a linear model between the observed data (i.e., the RF channel data) and the desired image to be recovered. Beamforming is then performed by solving a regularized inverse problem for each image depth separately. The results have been presented for focused~\cite{7565515} as well as plane-wave imaging~\cite{7728907}. Subsequently, this idea was further extended by considering more regularizations in the objective function and reconstructing all image depths jointly~\cite{8052532}. Two matrix-free formulations have been proposed in~\cite{8091286} that are both faster and more memory efficient than other inverse problem formulations. Recently, denoising-based regularization terms have been adapted in the inverse problem of ultrasound beamforming~\cite{9856693}, which noticeably improves the contrast and preserves the speckle statistics.\par
The last group of ultrasound beamforming methods is based on deep learning~\cite{8302520,8663450,9475029,9178454}. While deep models have great potential for estimating non-linear mapping functions between high dimensional input-output pairs and solving ill-posed problems, deep beamformers are subject to the following limitations. First, deep learning requires a massive amount of training data which is commonly unavailable. Second, the training ground truth is not known specifically for \textit{in vivo} data. Third, there is a noticeable performance reduction on test data due to domain shift between training and test data. The challenge on ultrasound beamforming with deep learning (CUBDL) was organized in conjunction with the 2020 IEEE International Ultrasonics Symposium (IUS)~\cite{9475029}. In~\cite{9251565}, a general ultrasound beamformer was designed, based on deep learning, to mimic MVB. In terms of image quality, it was ranked first. Overall, considering the network size as well, it was jointly ranked first with another submission~\cite{9251322}.\par
Finally, in ultrasound image deconvolution, Taxt~\textit{et al.} proposed a 2-D blind homomorphic approach wherein the PSF is estimated in the complex cepstrum domain followed by Wiener filtering for the deconvolution~\cite{935701}. An approach based on parametric inverse filtering was proposed in~\cite{4376242}. Subsequently, Yu~\textit{et al.} introduced a single-input multiple-output (SIMO) channel model for deconvolution of ultrasound images~\cite{6156829}. Two frameworks of Compressive Sensing (CS) and deconvolution were combined in~\cite{7302609} and the resulting method is called compressive deconvolution. An analytical model for the spatially-varying PSF in ultrasound imaging is proposed in~\cite{8092301}. A physical model for the nonstationary blur in plane- and diverging-wave imaging is proposed in~\cite{8636261}.\\*
Recently, the nonlinearity of ultrasound wave propagation in the tissue was considered in the deconvolution problem, and the enhanced image was reconstructed by the minimization of a joint cost function including the deconvolution models for both fundamental and harmonic RF images~\cite{9210758}. Most of the deconvolution studies reviewed here are categorized as blind methods as they are based on the estimation of PSF of the imaging system, while non-blind methods~\cite{393097} assume that the PSF is known (e.g., through experimental measurement).
%%%%%%%%%%%%%%%%%%%%%%%%%%%%%%%%%%%%%%%%%%%%%%%%%%%%%%%%%%%%%%%%%%%%%%%%%%%%%%%%%%%%%%%%%%%%%%%%%%%%%%%%
\section{Background}
\label{sec:sec2}
\subsection{Inverse problem of ultrasound beamforming}
\label{sec:sec21}
\begin{figure}[t]
	\captionsetup{justification=centering}
	\centerline{\includegraphics[width=8.5cm]{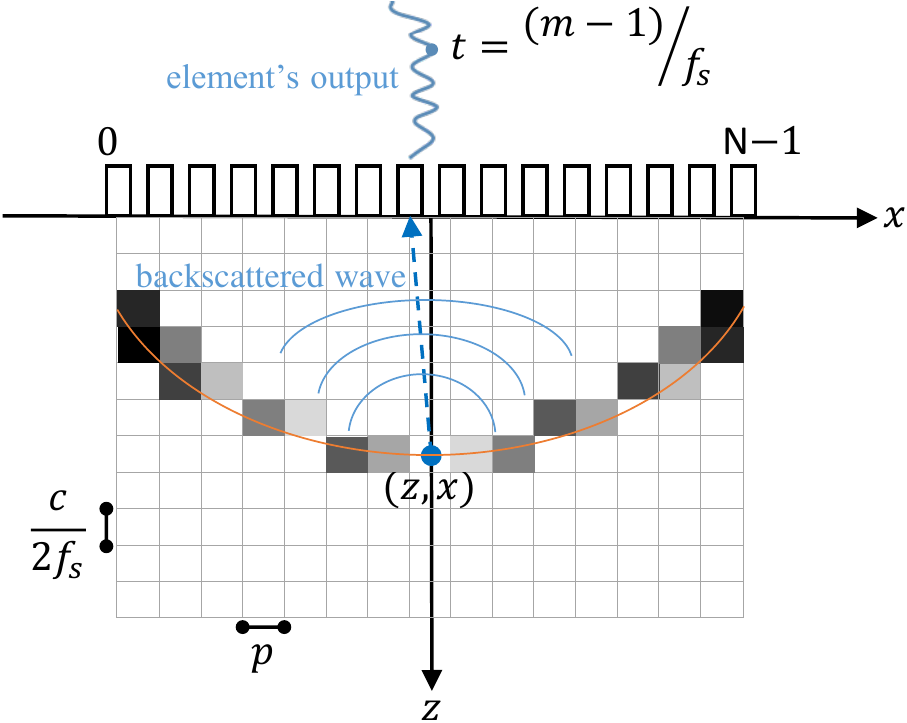}}
	\caption{The illustration of image pixels' contribution into a single sample of pre-beamformed data (adapted from~\cite{9856693}).}
	\label{fig:fig5}
\end{figure}

The purpose of ultrasound beamforming is to reconstruct a high-quality spatial map of the medium echogenicity. Without loss of generality, let us consider a N-element ultrasound linear probe with a transducer pitch of $p$, as shown in Fig.~\ref{fig:fig5}, from which $L$ piezoelectric elements transmit an acoustic wave into a medium with the constant sound speed of $c$. We also assume that backscattered signals are recorded with all elements of the same probe with a specific sampling frequency ($f_s$). To form a single image, this process may be repeated several times depending on the probe type (e.g., linear, phased array, or curvilinear) and the imaging technique (e.g., plane-wave, line-per-line, or synthetic aperture imaging). The beamforming grid is partitioned with a pixel sizes of $d_z = \frac{c}{2f_s}$ in the axial (i.e., the wave propagation direction) and $d_x = p$ in the lateral directions.\par
If there is no time offset after wave transmission (i.e., the elements immediately start to collect the backscattered waves), $m^{th}$ sample of the elements' outputs corresponds to the actual time of $t=(m-1)/f_s$, where $m=\{1, 2, ..., M\}$. In order to trace back the echoes corresponding to each
pixel, the associated time delay ($\tau$) equal to the sum of two-way propagation times of transmitted wave reaching that pixel ($\tau_t$) and getting back to the transducer elements ($\tau_r$) needs to be applied to each signal recorded by piezoelectric elements of the probe. Considering digitization error, the following condition determines all pixels contributing to a sample of element's output:
\begin{equation} 
\label{eq:1}
\mid t-\tau \mid \leq \frac{1}{f_s} ,
\end{equation}
where $\tau=\tau_t+\tau_r$ depends on pixel location, assumed speed of sound, the type of transmitted ultrasound wave (e.g., plane-wave, focused, or spherical wave), and the probe geometry. Eq.~(\ref{eq:1}) leads to an elliptical region with different weights as illustrated in Fig.~\ref{fig:fig5}. 
Therefore, a linear forward model between each sample of the RF channel data and the pixels' values in the desired image can be written as follows: 
\begin{equation} 
\label{eq:2}
\mathbf{y}_{ch} = \Phi \mathbf{x} + \mathbf{\nu} ,
\end{equation}
where $\mathbf{x}, \mathbf{y}_{ch}\in\mathbb{R}^{MN}$ are the vectorized versions of the desired image and the collected pre-beamformed channel data, respectively. $\Phi\in\mathbb{R}^{MN\times MN}$ is the weighting matrix, and $\mathbf{\nu}$ stands for the additive white Gaussian noise (AWGN).\par
The way matrix $\Phi$ is designed has been described in our previous work~\cite{9856693}. In this method, the linear propagation in the medium is assumed and PSF of the probe is not incorporated.
As Fig.~\ref{fig:fig5} and Eq.~(\ref{eq:2}) depict, rows of matrix $\Phi$ contain the contributions of image pixels into samples of RF channel data. For every single sample of pre-beamformed data, the pixels' contributions are determined using the following equation:
\begin{equation} 
\label{eq:3}
\Phi(i,j) = \begin{cases}1-\frac{\mid t_i-\tau_j \mid}{t_{max}} & \mid t_i-\tau_j \mid \leq \frac{1}{f_s}\\0 & \mid t_i-\tau_j \mid > \frac{1}{f_s}\end{cases},
\end{equation}
where $t_i$ is the actual time corresponding to a sample of element's output, and $\tau_j$ is the propagation delays of pixels which contribute to that sample (i.e., only the ones respecting the condition of Eq.~(\ref{eq:1})). $t_{max}$ is the maximum absolute difference between $t_i$ and $\tau_j$. Since only a small portion of pixels satisfy Eq.~(\ref{eq:1}), matrix $\Phi$ becomes highly sparse. Furthermore, data independent matrix $\Phi$ can be precalculated based on the known imaging settings. A reception apodization matrix, commonly used in DAS beamforming, is also multiplied with matrix $\Phi$ in order to take into account the directionality of transducer elements and fix the f-number for the entire image depths. It has to be mentioned that matrix $\Phi$ is not necessarily a square matrix because it can be determined for any grid partitioning not equal to pre-beamformed data. More details regarding the construction of matrix $\Phi$ can be found in~\cite{9856693}.\par 
The most straightforward inverse problem formulation of ultrasound beamforming is to estimate $\mathbf{x}$ by solving the following least-squares optimization problem:
\begin{equation} 
\label{eq:4}
\hat{\mathbf{x}} = \operatorname*{argmin}_{\mathbf{x}}  \parallel \mathbf{y}_{ch} - \Phi \mathbf{x} \parallel_2^2  .
\end{equation}
In contrast to common beamforming approaches (such as DAS, MVB, CF, etc), there is no need to estimate the apodization weights, and the beamformed image is directly reconstructed. It has been shown in~\cite{7565515,8052532} that Eq.~(\ref{eq:4}) does not solely provide the best results, and additional constrains should be considered. Furthermore, Eq.~(\ref{eq:4}) can be sequentially solved for each depth of the image (as performed in~\cite{7565515}) or solved at once for the whole image depths (as performed in~\cite{8052532}). Herein, we follow the second approach because of two main reasons. First, independently solving the inverse problem for each depth increases the computational cost of the algorithm noticeably. Second, we have observed border artifacts on the resulting image when different depths are reconstructed separately. 
%%%%%%%%%%%%%%%%%%%%%%%%%%%%%%%%%%%%%%%%%%%%%%%%%%%%%%%%%%%%%%%%%%%%%%%%%%%%%%%%%%%%%%%%%%%%%%%%%%%%%%%%
\subsection{Deconvolution of ultrasound images}
\label{sec:sec22}
Under the assumption of weak scattering for soft tissues and using the first-order Born approximation, the linear model, given in Eq.~(\ref{eq:2}), can also be used to express beamformed RF image as the result of convolution between the TRF and the PSF of the ultrasound imaging system as following~\cite{7302609,1539261,jensen1993}:
\begin{equation} 
\label{eq:6}
\mathbf{y}_{DAS} = H \mathbf{x} + \mathbf{\nu} ,
\end{equation}
where $\mathbf{y}_{DAS}$ is the RF image resulting from DAS beamforming. $H$ is a block circulant with circulant block (BCCB) matrix formed based on the PSF and accouting for circulant convolution. Although the assumption of the convolution model may not be valid in practice (especially for biological tissues), it has long been shown that the linear model is a good approximation and helps to reduce the adverse effect of PSF through deconvolution. Same as before (i.e., Eq.~(\ref{eq:4})), we consider the inverse problem formulation of deconvolution for finding the desired TRF.\par
The PSF is usually spatially variant in the ultrasound images mainly due to wave divergence, attenuation, and a limited number of crystal elements in the lateral direction. There are a few settings such as time gain compensation (TGC) and transmitting several focused beams (in line-per-line imaging) that help to have less variation in PSF across the image~\cite{9178454}. Therefore, a part of deconvolution studies solves the problem with the assumption of having a spatially-invariant PSF~\cite{7302609,4376242,9210758}. Herein, the experiments are based on plane-wave imaging and we have also considered a fixed PSF in our formulation.  
%%%%%%%%%%%%%%%%%%%%%%%%%%%%%%%%%%%%%%%%%%%%%%%%%%%%%%%%%%%%%%%%%%%%%%%%%%%%%%%%%%%%%%%%%%%%%%%%%%%%%%%%
\subsection{Basics of ADMM}
\label{sec:sec23}
As we use ADMM for solving the proposed optimization problem, a short overview of the method is provided in this section. More details of the ADMM algorithm can be found in~\cite{2200000016}.\par
Let us assume that our goal is to solve the following constrained optimization problem:
\begin{equation} 
\label{eq:7}
(\hat{\mathbf{u}},\hat{\mathbf{v}}) = \operatorname*{argmin}_{(\mathbf{u},\mathbf{v})}\; \{f(\mathbf{u})+g(\mathbf{v})\}\;\; s.t. \;\; A\mathbf{u}+B\mathbf{v}=\mathbf{c},
\end{equation}
where $\mathbf{u}, \mathbf{v}\in\mathbb{R}^n$, and $f:\mathbb{R}^n\rightarrow\mathbb{R}$ and $g:\mathbb{R}^n\rightarrow\mathbb{R}$ are closed convex functions. $A$ and $B$ are known matrices, and $\mathbf{c}$ is a given constant vector. To solve the corresponding unconstrained problem, the augmented Lagrangian function can be written as:
\begin{multline}
\label{eq:8}
\mathcal{L}(\mathbf{u},\mathbf{v},\mathbf{\lambda}) = f(\mathbf{u})+g(\mathbf{v})-\mathbf{\lambda}^T(A\mathbf{u}+B\mathbf{v}-\mathbf{c})\\+\frac{\beta }{2}\parallel A\mathbf{u}+B\mathbf{v}-\mathbf{c} \parallel_2^2 .
\end{multline}
The penalty term with parameter $\beta>0$ is added to enforce the constraint, and $\mathbf{\lambda}\in\mathbb{R}^n$ is the Lagrange multiplier.\par
Eq.~(\ref{eq:8}) can be written in an equivalent but more compact form as follows~\cite{Bouman2013}:
\begin{equation} 
\label{eq:9}
\mathcal{L}(\mathbf{u},\mathbf{v},\mathbf{\lambda}) = f(\mathbf{u})+g(\mathbf{v})+\frac{\beta }{2}\parallel A\mathbf{u}+B\mathbf{v}-\mathbf{c}+\frac{\mathbf{\lambda}}{\beta} \parallel_2^2 .
\end{equation}
The standard split-variable ADMM algorithm finds the solution of Eq.~(\ref{eq:9}) through an iterative process as following~\cite{2200000016}:
\begin{equation} 
\label{eq:10}
\left\{ \begin{array}{lcr}
\mathbf{u}^{i+1}=\operatorname{argmin}_{\mathbf{u}}\mathcal{L}(\mathbf{u},\mathbf{v}^i,\mathbf{\lambda}^i)\\ \mathbf{v}^{i+1}=\operatorname{argmin}_{\mathbf{v}}\mathcal{L}(\mathbf{u}^{i+1},\mathbf{v},\mathbf{\lambda}^i) \\ \mathbf{\lambda}^{i+1}=\mathbf{\lambda}^{i}+\beta(A\mathbf{u}^{i+1}+B\mathbf{v}^{i+1}-\mathbf{c})
\end{array}\right..
\end{equation}
As used in the next section, split-variable ADMM minimizes each term of the cost function separately. This property is beneficial in practice when a single optimization approach is not appropriate for both $f$ and $g$ functions. While the alternating minimization of $f$ and $g$ is much easier, it has been proven that ADMM iterations converge in convex optimization problems~\cite{2200000016}.
%%%%%%%%%%%%%%%%%%%%%%%%%%%%%%%%%%%%%%%%%%%%%%%%%%%%%%%%%%%%%%%%%%%%%%%%%%%%%%%%%%%%%%%%%%%%%%%%%%%%%%%%
\section{Proposed joint beamforming-deconvolution algorithm}
\label{sec:sec3}
The main idea of the current work is to find the desired TRF by solving a joint inverse problem of beamforming and deconvolution. Using the same variables introduced in Section~\ref{sec:sec2}, the proposed optimization problem is as follows:
\begin{multline}
\label{eq:11}
\hat{\mathbf{x}} = \operatorname*{argmin}_{\mathbf{x}} \frac{\gamma_D}{2} \parallel \mathbf{y}_{DAS} - H \mathbf{x} \parallel_2^2\\ + \frac{\gamma_B}{2} \parallel \mathbf{y}_{ch} - \Phi \mathbf{x} \parallel_2^2+\mu\parallel\mathbf{x}\parallel_1 .
\end{multline}
In addition to least-square terms for beamforming and deconvolution, the $\ell_1$-norm regularization term is also considered to enforce the sparsity of the solution, which is a common choice in ultrasound imaging~\cite{7565515,7728907,8052532}, but other regularization terms such as wavelet frames~\cite{8091286}, $\ell_2$-norm~\cite{7565515}, $\ell_p$-norms~\cite{7174535} can also be used with the proposed framework. $\gamma_D$, $\gamma_B$, and $\mu$ are  constant hyperparameters controlling the contribution of the deconvolution, beamforming, and sparse regularization terms, respectively. It is obvious that the objective function in Eq.~(\ref{eq:11}) is convex. The $L_1$ term, however, makes it nondifferentiable without a closed-form solution. Therefore, we split the independent variable $\mathbf{x}$ into three equivalent variables $\mathbf{u}$, $\mathbf{z}$, and $\mathbf{w}$ and consider the equality constraints. Hence, the new, but equivalent, form of Eq.~(\ref{eq:11}) is as following:
\begin{multline}
\label{eq:12}
(\hat{\mathbf{u}},\hat{\mathbf{w}},\hat{\mathbf{z}}) = \operatorname*{argmin}_{(\mathbf{u},\mathbf{w},\mathbf{z})} \frac{\gamma_D}{2} \parallel \mathbf{y}_{DAS} - H \mathbf{u} \parallel_2^2\\ + \frac{\gamma_B}{2} \parallel \mathbf{y}_{ch} - \Phi \mathbf{z} \parallel_2^2+\mu\parallel\mathbf{w}\parallel_1 s.t. \left\{ \begin{array}{lcr}
\mathbf{u} = \mathbf{z}\\ \mathbf{u} = \mathbf{w}
\end{array}\right..
\end{multline}
By looking closely at Eq.~(\ref{eq:12}), it can be considered as a specific form of the general formulation presented in Eq.~(\ref{eq:7}) with the following correspondences:
\begin{equation} 
\label{eq:13}
\left\{ \begin{array}{lcr}
f(\mathbf{u}) = \frac{\gamma_D}{2} \parallel \mathbf{y}_{DAS} - H \mathbf{u} \parallel_2^2\\ g(\mathbf{v}) = \frac{\gamma_B}{2} \parallel \mathbf{y}_{ch} - \Phi \mathbf{z} \parallel_2^2+\mu\parallel\mathbf{w}\parallel_1\\
\mathbf{v}=\left[ \begin{array}{c} \mathbf{w} \\ \mathbf{z} \end{array} \right],\mathbf{\lambda}=\left[ \begin{array}{c} \mathbf{\lambda_1} \\ \mathbf{\lambda_2} \end{array} \right]\\A=\left[ \begin{array}{c} I \\ I \end{array} \right],B=\left[ \begin{array}{cc} -I&0 \\ 0&-I \end{array} \right],\mathbf{c}=0
\end{array}\right.,
\end{equation} 
where $I$ refers to an identity square matrix of size $n$, therefore, the augmented Lagrangian function of Eq.~(\ref{eq:12}) is exactly the same as what was presented previously in Eq.~(\ref{eq:9}), and its solution can be found using the split-variable ADMM approach.\par
Algorithm~\ref{alg:1} describes the proposed solution of our cost function in the ADMM framework. Different terms of the Eq.~(\ref{eq:11}) are minimized separately in each iteration. The algorithm is initialized by setting the hyperparameters $\gamma_D$, $\gamma_B$, $\mu$, and $\beta$. A small constant value $\epsilon$ is chosen as the threshold for the stopping criterion. The iterative optimization procedure is started with arbitrary initial values for the Lagrange multiplier ($\lambda$) and the new variables (i.e., $\mathbf{u}$, $\mathbf{w}$, and $\mathbf{z}$). The proposed ADMM solution can be summarized in three steps as follows.
\begin{algorithm}[t]
	\caption{ADMM Algorithm for solving Eq.~(\ref{eq:11})}	
	\begin{algorithmic}[1]
		\label{alg:1}
		\STATE \textbf{Input:}\:$H$, $\Phi$, $\mathbf{y}$
		\STATE \textbf{Set:\:}$\gamma_D>0$, $\gamma_B>0$, $\mu>0$, $\beta>0$, $\mathbf{u}^{0}$, $\mathbf{v}^{0}$, $\mathbf{\lambda}^{0}$, $\epsilon$
		\STATE \textbf{While\:\:}stopping criterion\:$>\epsilon$\:\:\textbf{do}
		\STATE $\mathbf{u}^{i+1}=\operatorname{argmin}_{\mathbf{u}}\frac{\gamma_D}{2} \parallel \mathbf{y}_{DAS} - H \mathbf{u} \parallel_2^2+\frac{\beta }{2}\parallel A\mathbf{u}+B\mathbf{v}^i+\frac{\mathbf{\lambda}^i}{\beta} \parallel_2^2$
		\STATE $\mathbf{z}^{i+1}=\operatorname{argmin}_{\mathbf{z}}\frac{\gamma_B}{2} \parallel \mathbf{y}_{ch} - \Phi \mathbf{z} \parallel_2^2+\frac{\beta }{2}\parallel \mathbf{u}^{i+1}-\mathbf{z}+\frac{\mathbf{\lambda}_2^i}{\beta} \parallel_2^2$
		\STATE $\mathbf{w}^{i+1}=\operatorname{argmin}_{\mathbf{w}}\mu\parallel\mathbf{w}\parallel_1+\frac{\beta }{2}\parallel \mathbf{u}^{i+1}-\mathbf{w}+\frac{\mathbf{\lambda}_1^i}{\beta} \parallel_2^2$
		\STATE $\mathbf{\lambda}^{i+1}= \mathbf{\lambda}^i+\beta(A\mathbf{u}^{i+1}+B\mathbf{v}^{i+1})$
		\STATE \textbf{End}
	\end{algorithmic}
\end{algorithm}
%%%%%%%%%%%%%%%%%%%%%%%%%%%%%%%%%%%%%%%%%%%%%%%%%%%%%%%%%%%%%%%%%%%%%%%%%%%%%%%%%%%%%%%%%%%%%%%%%%%%%%%%
\subsection{Deconvolution update}
\label{sec:sec31}
In this step, the solution of the deconvolution term ($\mathbf{u}$) is found by minimizing the corresponding subproblem written in line 4 of Algorithm~\ref{alg:1}. Since this cost function is convex and differentiable, the solution can be easily found by taking the gradient and setting it to zero. By doing so, the following closed-form solution is derived:
\begin{multline}
\label{eq:14}
\mathbf{u}^{i+1}=\\({\gamma_D\,H^TH+2\beta J})^{-1}(\gamma_D\,H^T\mathbf{y}_{DAS}+\beta\mathbf{w}^i+\beta\mathbf{z}^i-\mathbf{\lambda}^i_1-\mathbf{\lambda}^i_2) ,
\end{multline}
where $J$ is a matrix of ones with the same size as $H^TH$. Eq.~(\ref{eq:14}) can also be solved in the Fourier domain. This implementation has been successfully used in~\cite{9210758,7302609} to reduce the computational complexity of the solution in each iteration.
%%%%%%%%%%%%%%%%%%%%%%%%%%%%%%%%%%%%%%%%%%%%%%%%%%%%%%%%%%%%%%%%%%%%%%%%%%%%%%%%%%%%%%%%%%%%%%%%%%%%%%%%
\subsection{Beamforming update}
\label{sec:sec32}
\begin{figure*}[t!]
	\centering
	\centerline{\includegraphics[width=\textwidth]{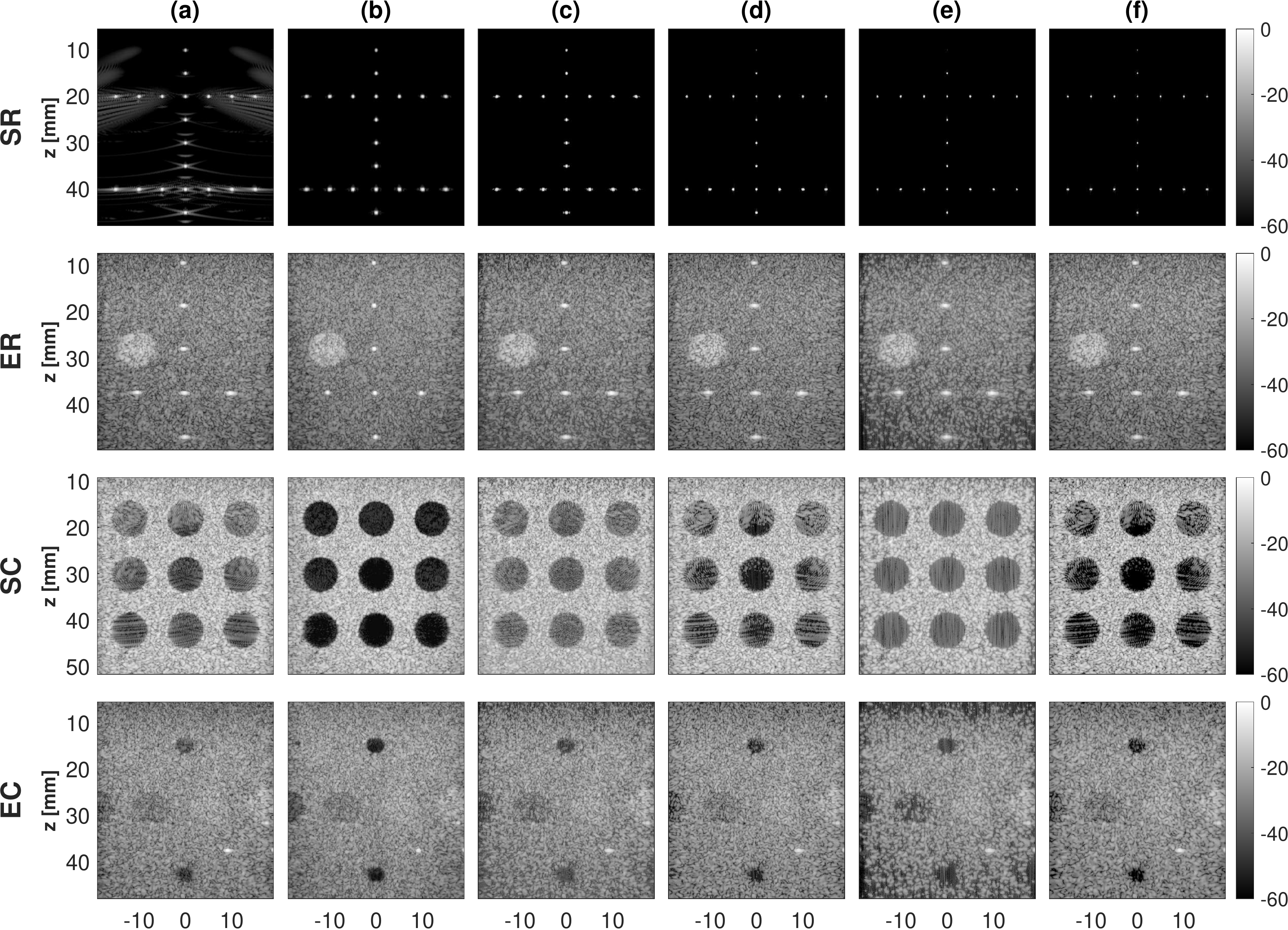}}
	%  \vspace{2.0cm}
	\caption{Simulation and experimental images reconstructed through different methods. Rows indicate datasets while columns correspond to different approaches. (a) DAS. (b) CPWC. (c) The inverse problem of beamforming. (d) The inverse problem of deconvolution. (e) Sequential approach. (f) The proposed joint formulation. CPWC is obtained from 75 steered insonifications. All other results are from a single $0^o$ insonification.}
	\label{fig:fig1}
\end{figure*}
The second step corresponds to the minimization of beamforming term written in line 5 of Algorithm~\ref{alg:1}. Same as step 1, if we set the gradient of the cost function to zero, we arrive at the following analytical solution:
\begin{equation}
\label{eq:15}
\mathbf{z}^{i+1}=(\gamma_B\Phi^T\Phi+\beta J)^{-1}(\gamma_B\,\Phi^T\mathbf{y}_{ch}+\beta\mathbf{u}^{i+1}+\mathbf{\lambda}^i_2) ,
\end{equation}
where $\Phi^T\Phi$ is usually a square matrix of size several hundreds of thousands, and not even diagonal, nor one that can be diagonalizable through Fourier transform. Therefore, Eq.~(\ref{eq:15}) is intractable in practice as involves large matrix inversion. That is why a numerical method is adopted to tackle this problem and find the solution of this step. Herein, the optimal solution is found using limited-memory Broyden–Fletcher–Goldfarb–Shanno (BFGS) solver\footnote{MATLAB implementation is publicly available in this link:~\url{http://www.cs.ubc.ca/~schmidtm/Software/minFunc.html}}, which is a quasi-Newton approach. Limited-memory BFGS achieves quadratic convergence for many problems~\cite{Nocedal1} and is highly efficient for unconstrained optimization of differentiable real-valued high-dimensional functions. It has to be emphasized that the modular property of ADMM provides the possibility of solving the beamforming term using limited-memory BFGS in which the Hessian matrix is approximated. This property is worth noting in our problem wherein the large size of matrix $\Phi$ makes the calculation of Hessian intractable.
%%%%%%%%%%%%%%%%%%%%%%%%%%%%%%%%%%%%%%%%%%%%%%%%%%%%%%%%%%%%%%%%%%%%%%%%%%%%%%%%%%%%%%%%%%%%%%%%%%%%%%%%
\subsection{Sparsity and Lagrange multiplier updates}
\label{sec:sec33}
The final step of our method entails the optimization of sparsity constraints and updating the Lagrangian multiplier. The minimization problem presented in line 6 of Algorithm~\ref{alg:1} is commonly called as the proximal mapping of the $L_1$ norm as follows~\cite{Bouman2013}:
\begin{multline}
\label{eq:16}
prox_{\mu\parallel.\parallel_1/\beta}(\mathbf{u}^{i+1}+\frac{\mathbf{\lambda}_1^i}{\beta})=\operatorname{argmin}_{\mathbf{w}}\mu\parallel\mathbf{w}\parallel_1+\\\frac{\beta }{2}\parallel \mathbf{u}^{i+1}-\mathbf{w}+\frac{\mathbf{\lambda}_1^i}{\beta} \parallel_2^2 .
\end{multline}
Eq.~(\ref{eq:16}) is the minimization of a strictly convex function, and its unique minimizer can be presented in terms of shrinkage function~\cite{Bouman2013}, which acts as a soft-thresholding operator:
\begin{multline}
\label{eq:17}
\mathbf{w}^{i+1}=soft_{\mu/\beta}(\mathbf{u}^{i+1}+\frac{\mathbf{\lambda}_1^i}{\beta})=\\=max\{|\mathbf{u}^{i+1}+\frac{\mathbf{\lambda}_1^i}{\beta}|-\frac{\mu}{\beta},0\}sign(\mathbf{u}^{i+1}+\frac{\mathbf{\lambda}_1^i}{\beta}) .
\end{multline}
Finally, the Lagrangian multiplier needs to be updated using the equation in line 7 of Algorithm~\ref{alg:1}.\par
In each iteration of Algorithm~\ref{alg:1}, the original objective function (i.e., Eq.~(\ref{eq:11})) is calculated, and its relative error for two consecutive iterations is used as the stopping criterion. The theoretical convergence of the split variable ADMM algorithm to a global minimum in any convex optimization problem has been shown~\cite{eckstein2015}.  
%%%%%%%%%%%%%%%%%%%%%%%%%%%%%%%%%%%%%%%%%%%%%%%%%%%%%%%%%%%%%%%%%%%%%%%%%%%%%%%%%%%%%%%%%%%%%%%%%%%%%%%%
\section{Experiments}
\label{sec:sec4}
\subsection{Dataset}
\label{sec:sec41} 
The performance of the proposed method is evaluated on a set of simulations, real phantoms, and \textit{in vivo} publicly available PICMUS\footnote{The datasets are publicly available at PICMUS website:~\url{https://www.creatis.insa-lyon.fr/Challenge/IEEE IUS 2016/}} benchmark datasets~\cite{7728908}. PICMUS is the first beamforming competition that was held in conjunction with the 2016 IEEE International Ultrasonics Symposium (IUS) and designed to create a benchmark plane-wave dataset and facilitate the comparison of different approaches~\cite{7728908}.
Details regarding PICMUS datasets can be found in~\cite{7728908} and are not repeated here to keep the paper concise. In short, simulation and experimental phantom images containing point targets are considered to assess the proposed methods in terms of spatial resolution (denoted by Simulation Resolution (SR) and Experimental Resolution (ER)). Also, simulation and experimental phantom images containing anechoic cysts are considered to assess the proposed methods in terms of contrast (denoted by Simulation Contrast (SC) and Experimental Contrast (EC)). \textit{In vivo} images have been collected from the carotid artery of a volunteer subject, including cross-sectional (denoted by Carotid Cross (CC)) and longitudinal views (denoted by Carotid Longitudinal(CL)).
%%%%%%%%%%%%%%%%%%%%%%%%%%%%%%%%%%%%%%%%%%%%%%%%%%%%%%%%%%%%%%%%%%%%%%%%%%%%%%%%%%%%%%%%%%%%%%%%%%%%%%%%
\subsection{Evaluation metrics}
\label{sec:sec42}
The images reconstructed using the proposed method are evaluated in terms of two main specialized ultrasound assessment indexes, including resolution and contrast.\par
As for the resolution index, the Full Width at Half Maximum (FWHM) is calculated in the axial and lateral directions. The contrast index is reported using two different criteria, and the Contrast-to-Noise Ratio (CNR) is calculated as follows:
\begin{equation} 
\label{eq:18}
CNR=20log_{10}(\frac{ \mid\mu_{ROI}-\mu_{B}\mid}{ \sqrt{(\sigma_{ROI}^2+\sigma_{B}^2)/2}}),
\end{equation} 
where $\sigma_{ROI}$ and $\sigma_{B}$ are the standard deviation of the image over the region of interest (ROI) and background, respectively. $ \mu_{ROI}$ and $\mu_{B}$ are the means of image over the ROI and background, respectively.\par  
Recently, it has been shown that quantitative indexes such as CNR and FWHM are not reliable when the dynamic range of the final image has been transformed~\cite{9247962,9662310}. For example, FWHM improves by taking the square of an image while there is no new information content~\cite{9662310}. Therefore, it is recommended to use histogram matching (HM) prior to the visualization and making the quantitative measurements, which lead to fairer comparisons between different methods and would be an acceptable alternative~\cite{9247962,9662310}. In the current study, HM is applied to all the presented results in the next section except for the SR experiment for which HM brings additional artifacts due to its binary content. Since a homogeneous speckle region of interest (ROI) has a well-behaved log-Rayleigh distribution for B-mode images, ROI-based HM (details can be found in~\cite{9247962}) is applied to the result of each method, and DAS output is considered as the reference image.\par   
We have also reported generalized CNR (gCNR) criterion introduced in~\cite{8918059}, which has been shown to be robust against dynamic range alterations. gCNR formula is as follows:
\begin{equation} 
\label{eq:19}
gCNR = 1- \int_{-\infty}^{\infty} min\left\{p_{ROI}(x),p_B(x)\right\}dx ,
\end{equation} 
where $p_B(x)$ and $p_{ROI}(x)$ are the histograms of pixels measured in the background and ROI, respectively. gCNR specifies the overlap between intensity distributions of two regions regardless of grayscale intensity transformations. Higher distributions overlap leads to lower gCNR values. When the two distributions are independent, gCNR is equal to its maximum value of 1~\cite{8918059}.\par 
%%%%%%%%%%%%%%%%%%%%%%%%%%%%%%%%%%%%%%%%%%%%%%%%%%%%%%%%%%%%%%%%%%%%%%%%%%%%%%%%%%%%%%%%%%%%%%%%%%%%%%%%
\section{Results}
\label{sec:sec5}
\begin{table}[t!]
	\caption{Quantitative results in terms of resolution and contrast indexes for simulation and real phantom experiments.}
	\label{table:1}
	\centering
	\setlength{\tabcolsep}{2.5pt}
	\scriptsize
	\begin{tabular}{c c c c c c c c c c c}
		\specialrule{.15em}{0em}{.2em}
		dataset & SR & ER & SC & EC  \\ [.2em] 
		\specialrule{.05em}{0em}{.2em} 
		index & FWHM\textsubscript{A} FWHM\textsubscript{L} & FWHM\textsubscript{A} FWHM\textsubscript{L} & CNR gCNR & CNR gCNR \\ [.2em] 
		\specialrule{.05em}{0em}{.2em} 
		\makecell{DAS \\ CPWC \\ Beamforming \\ Deconvolution \\ Sequential \\ Joint} & \makecell{0.4 \\ 0.4 \\ 0.38 \\ 0.29 \\ 0.26 \\ 0.22} \makecell{0.47 \\ 0.4 \\ 0.4 \\ 0.34 \\ 0.27 \\ 0.26} &    \makecell{0.57 \\ 0.39 \\ 0.56 \\ 0.37 \\ 0.36 \\ 0.34} \makecell{0.96 \\ 0.39 \\ 0.94 \\ 0.55 \\ 0.5 \\ 0.46} & \makecell{10.25 \\ 17.57 \\ 10.12 \\ 9.92 \\ 10.04 \\ 11.54} \makecell{0.89 \\ 0.99 \\ 0.9 \\ 0.88 \\ 0.88 \\ 0.95}&  \makecell{7.75 \\ 12.9 \\ 7.65 \\ 8.15 \\ 8.1 \\ 9.15} \makecell{0.8 \\ 0.94 \\ 0.78 \\ 0.79 \\ 0.8 \\ 0.85}\\ [.2em] 
		\specialrule{.05em}{0em}{.2em}
		\makecell{EMV \\ PCF \\ Stolt \\ UFSB} & \makecell{0.4 \\ 0.3 \\ 0.42 \\ 0.4} \makecell{0.1 \\ 0.38 \\ 1.12 \\ 0.85} &    \makecell{0.51 \\ 0.46 \\ 0.3 \\ 0.44} \makecell{0.46 \\ 0.41 \\ 0.28 \\ 0.42} & \makecell{9.34 \\ 7.86 \\ 2.3 \\ 7.85} \makecell{0.85 \\ 0.79 \\ 0.55 \\ 0.79} &  \makecell{8.25 \\ 6 \\ 7.2 \\ 7} \makecell{0.84 \\ 0.7 \\ 0.79 \\ 0.78} \\ [.2em] 
		\specialrule{.05em}{0em}{.2em}
	\end{tabular}
\end{table}
In this section, the results of the proposed joint formulation are compared with the sequential approach that entails beamforming followed by deconvolution. To better understand the effect of each term in the proposed joint cost function, the images reconstructed by only solving the inverse problem  of beamforming (i.e., $\gamma_D = 0$ and $\gamma_B = 1$ in Eq.~(\ref{eq:11})) or deconvolution (i.e., $\gamma_D = 1$ and $\gamma_B = 0$in Eq.~(\ref{eq:11})) are also presented in Section~\ref{sec:sec51}.
Furthermore, to demonstrate the superiority of the proposed approach in comparison with the previous beamforming methods, the results of Eigenspace-based MV (EMV)~\cite{5611687}, PCF~\cite{4976281}, Fourier domain technique based on Stolt’s migration~\cite{8359331}, and ultrasound Fourier slice
beamforming (UFSB)~\cite{7582552} are included in Section~\ref{sec:sec52}.
Finally, the sensitivity analysis of the proposed method to initial points and parameter selection is presented in Section~\ref{sec:sec53}.\par
Hereafter,  we consider the Hanning apodization window with $f\# = 0.5$ for DAS and other methods on top of DAS (except for \textit{in vivo} datasets for which the Tukey (tapered cosine) window with constant parameter of 0.25 and f-number equals to 1.75 is considered). $\epsilon = 10^{-3}$ is selected as the threshold for stopping criterion. The iterative algorithms are started with initial values equal to zero. Since the proposed joint formulation and also deconvolution approach require PSF, the method proposed in~\cite{1375163} has been adopted to estimate the unknown PSF from the RF data. The quantitative indexes are calculated independently for different point targets or cyst regions, and the average values are reported. The hyperparameters of each method are set independently to achieve the best results. Therefore, detailed information on the selected hyperparameters of each method is provided in the Appendix Section.
%%%%%%%%%%%%%%%%%%%%%%%%%%%%%%%%%%%%%%%%%%%%%%%%%%%%%%%%%%%%%%%%%%%%%%%%%%%%%%%%%%%%%%%%%%%%%%%%%%%%%%% 
\subsection{The proposed joint formulation}
\label{sec:sec51}
\begin{figure*}[t!]
	\centering
	\centerline{\includegraphics[width=\textwidth]{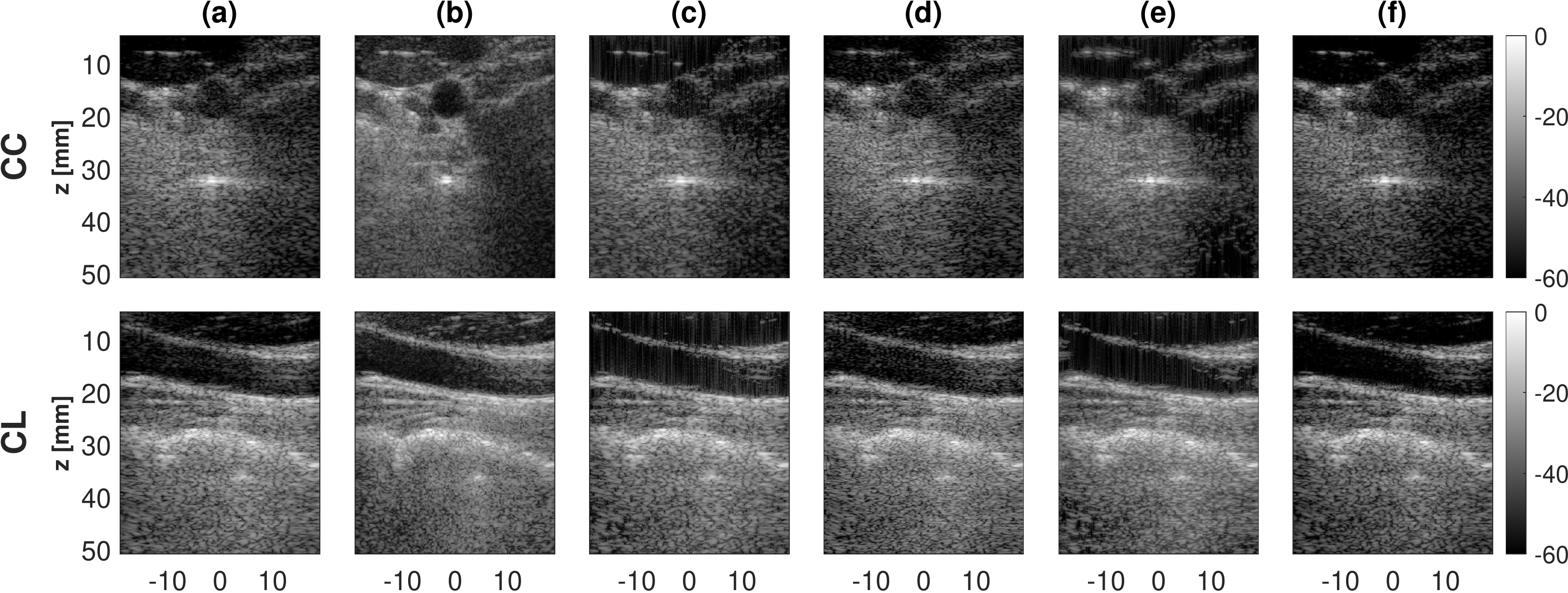}}
	%  \vspace{2.0cm}
	\caption{Results on \textit{in vivo} data. Rows indicate datasets while columns correspond to different approaches. (a) DAS. (b) CPWC. (c) The inverse problem of beamforming. (d) The inverse problem of deconvolution. (e) Sequential approach. (f) The proposed joint formulation. CPWC is obtained from 75 steered insonifications. All other results are from a single $0^o$ insonification.}
	\label{fig:fig2}
\end{figure*}
%%%%%%%%%%%%%%%%%%%%%%%%%%%%%%%%%%%%%%%%%%%%%%%%%%%%%%%%%%%%%%%%%%%%%%%%%%%%%%%%%%%%%%%%%%%%%%%%%%%%%%%%
\subsubsection{Simulation and experimental data}
\label{sec:sec511}
The images reconstructed from the single $0^o$ plane-wave are presented in Fig.~\ref{fig:fig1}. As for the reference quality, the results of Coherent Plane-Wave Compounding (CPWC) on 75 insonifications are illustrated in the second column of Fig.~\ref{fig:fig1}. The proposed joint formulation can successfully reconstruct high-quality images in terms of both resolution and contrast. Fig.~\ref{fig:fig1} shows that the sequential approach has a poor performance because a part of image content get lost in the two consecutive soft-thresholding steps. As for the result of beamforming and deconvolution alone, the artificial improvement in contrast has been revoked by HM. And Fig.~\ref{fig:fig1} depicts that only the results of CPWC and the proposed method are robust to HM.\par   
As quantitative results of Table~\ref{table:1} confirm, the proposed method gives high axial and lateral resolutions for both simulation and experimental data similar to or even better than CPWC. The highest contrast corresponds to CPWC results since it averages over 75 angles and perfectly suppresses the side-lobe artifacts. For a single $0^o$ insonification, the proposed approach improves the contrast as compared to other methods. This point can also be seen in the quantitative comparison reported in Table~\ref{table:1}.\par
%%%%%%%%%%%%%%%%%%%%%%%%%%%%%%%%%%%%%%%%%%%%%%%%%%%%%%%%%%%%%%%%%%%%%%%%%%%%%%%%%%%%%%%%%%%%%%%%%%%%%%%%
\subsubsection{\textit{In vivo} data}
\label{sec:sec512}  
The proposed method is also evaluated on real data collected from the carotid artery. The visual comparison of different approaches is illustrated in Fig.~\ref{fig:fig2}. As can be seen in Fig.~\ref{fig:fig2}, the proposed method is able to suppress the clutter artifacts caused by diffuse reverberation from shallow layers and create a dark image of the artery in both cross-sectional and longitudinal views. 
%%%%%%%%%%%%%%%%%%%%%%%%%%%%%%%%%%%%%%%%%%%%%%%%%%%%%%%%%%%%%%%%%%%%%%%%%%%%%%%%%%%%%%%%%%%%%%%%%%%%%%%%
\subsection{Comparison with other methods}
\label{sec:sec52}
\begin{figure*}[t!]
	\centering
	\centerline{\includegraphics[width=0.7\textwidth]{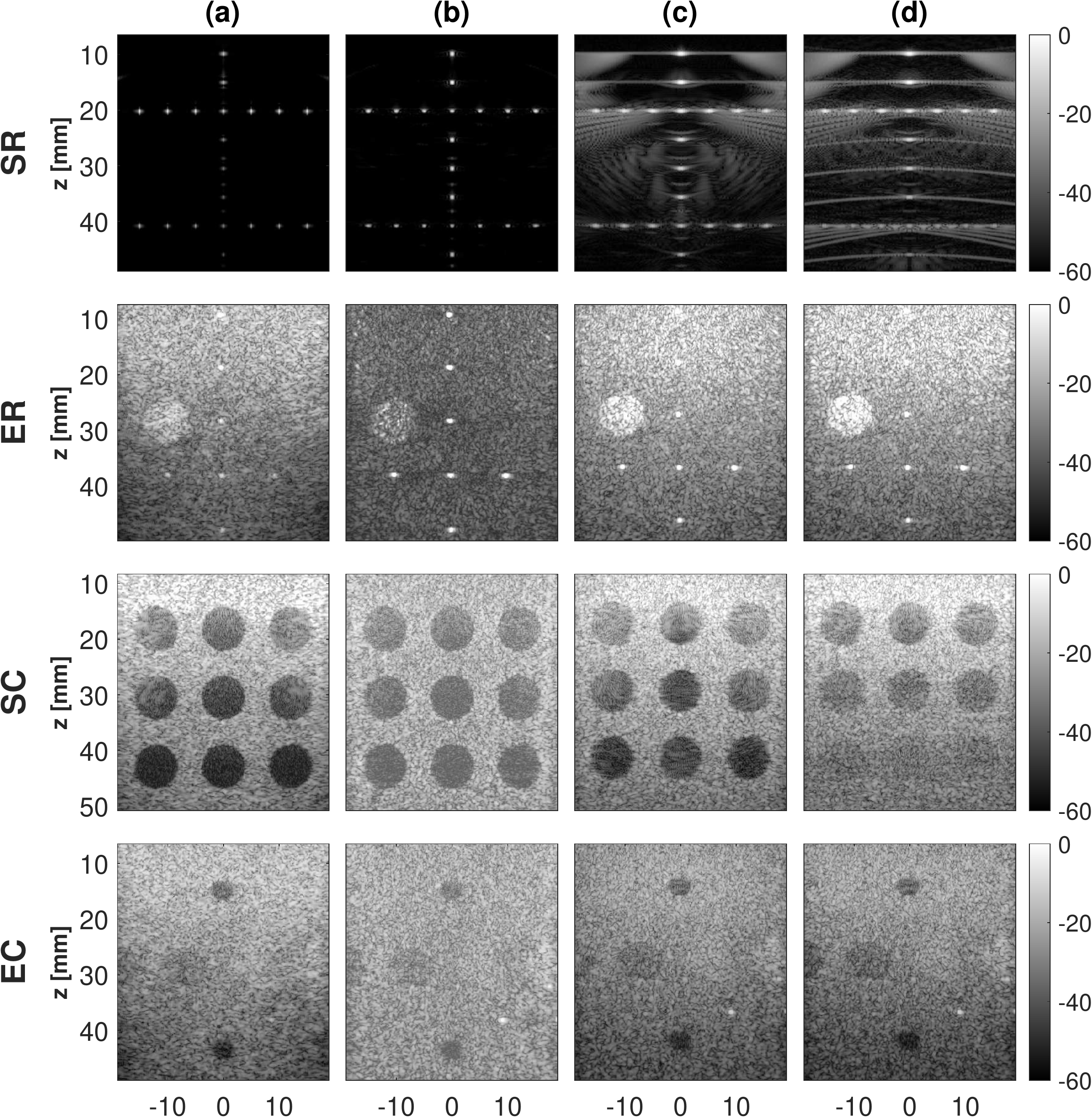}}
	%  \vspace{2.0cm}
	\caption{Simulation and experimental images reconstructed through previous beamforming approaches. Rows indicate datasets while columns correspond to different approaches. (a) EMV~\cite{5611687}. (b) PCF~\cite{4976281}. (c) UFSB~\cite{7582552}. (d) Stolt’s migration~\cite{8359331}. All the results are from a single $0^o$ insonification.}
	\label{fig:fig3}
\end{figure*}
In this section, the results of previous beamforming methods are shown. We have included an example from each group of classical beamforming methods reviewed in Section~\ref{sec:sec11}. The results of DAS and inverse problem formulation are presented along with the proposed approach in Section~\ref{sec:sec51}. The visual comparison of results, presented in Fig.~\ref{fig:fig3}, indicates that EMV only gives a better lateral resolution for SR dataset as compared to the proposed method. This point can also be understood from the quantitative comparison reported in Table~\ref{table:1}. PCF works better than DAS in terms of axial resolution, while its results are still worse than the proposed method. In terms of lateral resolution, however, the results are discussible. While EMV substantially improves the lateral resolution of simulation data, the same improvement was not achieved in the real phantom experiment. The reason behind this difference is the number of eigenvectors of the covariance matrix used for creating the signal subspace. More specifically, EMV only needs the principal eigenvectors of the SR data to successfully reconstruct the image because it only includes some point targets. However, if we only consider some of the eigenvectors in the ER case, a part of the speckle texture would be lost. The same difference in improvement can be noticed in terms of contrast. That is because of the additional noise of experimental data, which reduces the quality of covariance matrix estimation.\par
In short, the comparison with previous approaches reveals that the proposed method gives the most consistent improvement in image quality for all datasets. Although other methods may provide better qualities in specific cases, their performance drops in other experiments. This point can also be seen for Stolt’s migration~\cite{8359331} approach, which gives a high resolution in ER experiment but has a poor performance for other datasets.   
%%%%%%%%%%%%%%%%%%%%%%%%%%%%%%%%%%%%%%%%%%%%%%%%%%%%%%%%%%%%%%%%%%%%%%%%%%%%%%%%%%%%%%%%%%%%%%%%%%%%%%%%
\subsection{Sensitivity analysis}
\label{sec:sec53}
As mentioned in Algorithm~\ref{alg:1}, the proposed method is initialized with the arbitrary values for the Lagrange multiplier ($\lambda$) and new variables (i.e., $\mathbf{u}$, $\mathbf{w}$, and $\mathbf{z}$). Furthermore, hyperparameters $\gamma_D$, $\gamma_B$, $\mu$, and $\beta$, which respectively specify the weights of deconvolution, beamforming, sparsity, and penalty terms, need to be set.\par
Generally, the initial values of $\lambda$ and new variables do not make any difference to the final solution because the proposed objective function is convex, and there is no local minimum. Notwithstanding, the starting points can affect the processing time and the number of iterations for convergence. For all the results presented here, zero is used as the initial value, and the algorithm always converges in less than 30 iterations. The code is implemented in Windows 10 using the MATLAB R2021a programming platform.\par
Different hyperparameters' values may completely change the final solution. Generally, a large $\beta$ forces the algorithm to perfectly accomplish the equality constraints of new variables (Eq.~(\ref{eq:12})) while a small $\beta$ allows the algorithm to converge toward different values for $\mathbf{u}$, $\mathbf{w}$, and $\mathbf{z}$. While deconvolution and beamforming terms have equal weights, a large $\mu$ may wipe out the speckle texture, and the resulting image looks too dark, and if we set $\mu=0$, the contrast of the resulting image would be poor. Finally, if any deconvolution or beamforming terms become dominant, the resulting quality will not be desirable.
%%%%%%%%%%%%%%%%%%%%%%%%%%%%%%%%%%%%%%%%%%%%%%%%%%%%%%%%%%%%%%%%%%%%%%%%%%%%%%%%%%%%%%%%%%%%%%%%%%%%%%%%
\section{Discussions}
\label{sec:sec6}
The sequential approach and joint formulation comparison confirm that solving each beamforming and deconvolution problem separately does not lead to the same quality. This might be due to a loss of information in the first step of the sequential approach. The same observation has been reported in~\cite{7302609}.\par
Another important advantage of the proposed formulation is the substantial improvement in axial resolution. As reported in Section~\ref{sec:sec52}, other beamforming approaches either do not change or have a lower effect on the axial resolution as compared to our results. This point is crucial because the resolution in the axial direction is usually increased by transmitting pulses with a higher center frequency.\par
The achieved improvement in image quality can reduce the need to transmit several plane-waves with different angles. Furthermore, the proposed framework provides the possibility of considering the result of any other beamforming approach (not only DAS) in the deconvolution term. In other words, the proposed objective function, in Eq.~(\ref{eq:11}), is a linear combination of beamforming and deconvolution terms. Therefore, extra terms based on the result of other beamforming approaches can also be added.\par  
Using ADMM makes the optimization step of the proposed method easy to implement and reduces computational costs. As the proposed algorithm is iterative and the variables are updated serially, parallel implementation is impossible. Therefore, although our method is much faster than a computationally expensive algorithm such as MV, real-time image reconstruction using the proposed method might not be possible. This would be the subject of our future research.\par
Although the proposed idea can be applied to any imaging technique (i.e., focused, plane-wave, and synthetic aperture imaging) or probe type (i.e., linear, convex, and phased array), only the benchmark PICMUS dataset is used here because it is publicly available and the comparison with previous approaches is easier. It also helps the readers to reimplement the algorithm and verify the results easily. The proposed method can also be applied on top of the CPWC. However, to limit the sources of improvement, CPWC results are not used.\par
It has been previously shown that solving the inverse problem of beamforming gives images with high resolution and contrast~\cite{7565515,8052532,7728907}. This improvement, however, comes at the expense of speckle information loss. Furthermore, the inverse problem of deconvolution cannot solely improve the image quality. Combining both terms in our proposed objective function helps to achieve a high resolution and contrast while the speckle texture is also preserved. This point can be seen in ER data. The proposed method's ability to preserve the speckle texture is of crucial importance in image computing applications such as speckle tracking and tissue classification. In addition, experienced radiologists often rely on the speckle pattern for diagnosis.\par
The performance of the proposed method directly depends on the quality of the estimated PSF of the imaging system. Herein, we utilize a common approach used in previous literature~\cite{4376242,7302609,9210758}. However, any method for PSF estimation can also be used to improve the results. We plan to extend our idea to consider a nonstationary PSF in the model and also take advantage of the harmonic components in RF data for image reconstruction.      
%%%%%%%%%%%%%%%%%%%%%%%%%%%%%%%%%%%%%%%%%%%%%%%%%%%%%%%%%%%%%%%%%%%%%%%%%%%%%%%%%%%%%%%%%%%%%%%%%%%%%%%%
\section{Conclusions}
\label{sec:sec7}
Beamforming and deconvolution have only been used separately in a sequential approach. Herein, we proposed a novel formulation for combining both methods. A regularized inverse problem including two linear models for beamforming and deconvolution plus additional sparsity constraint is solved using the ADMM algorithm. The proposed image reconstruction approach is a joint optimization problem that uses DAS results as an observation. The results show that the proposed iterative method gives ultrasound images with a high resolution and contrast. 
%%%%%%%%%%%%%%%%%%%%%%%%%%%%%%%%%%%%%%%%%%%%%%%%%%%%%%%%%%%%%%%%%%%%%%%%%%%%%%%%%%%%%%%%%%%%%%%%%%%%%%%%
\section{Acknowledgment}
\label{sec:sec8}
The authors would like to thank the organizers of the PICMUS challenge and the ultrasound toolbox for providing publicly available codes and data. We sincerely thank Teodora Szasz, Adrien Besson, and Mohammed Albulayli for making their MATLAB codes publicly available.
%%%%%%%%%%%%%%%%%%%%%%%%%%%%%%%%%%%%%%%%%%%%%%%%%%%%%%%%%%%%%%%%%%%%%%%%%%%%%%%%%%%%%%%%%%%%%%%%%%%%%%%%
\section{Appendix}
\label{sec:sec9}
Table~\ref{table:3} includes the value of hyperparameters selected for different approaches in each experiment. In the EMV method, the subarray size equals 64, the temporal averaging factor is set to 1.5, and the diagonal loading is 0.01. The signal subspace is created using all eigenvectors of the covariance matrix except for the SR experiment, for which only considering the largest 10\% is enough. 
\begin{table}[t!]
	\caption{The value of hyperparameters of each method in different experiments.}
	\label{table:3}
	\centering
	\setlength{\tabcolsep}{2.5pt}
	\scriptsize
	\begin{tabular}{c c c c c c c c c c c}
		\specialrule{.15em}{0em}{.2em}
		method & Beamforming & Deconvolution & Joint  \\ [.2em] 
		\specialrule{.05em}{0em}{.2em} 
		hyperparameter & $\mu$ \thickspace\thickspace\thickspace\thickspace\thickspace\thickspace\thickspace $\beta$ &\thickspace\thickspace\thickspace $\mu$ \thickspace\thickspace\thickspace\thickspace\thickspace\thickspace\thickspace $\beta$\thickspace\thickspace\thickspace &\thickspace\thickspace\thickspace $\gamma_D$ \thickspace\thickspace\thickspace$\gamma_B$ \thickspace\thickspace\thickspace\thickspace\thickspace\thickspace\thickspace\thickspace\thickspace $\beta$ \thickspace\thickspace\thickspace\thickspace\thickspace\thickspace\thickspace\thickspace\thickspace $\mu$ \thickspace\\ [.2em] 
		\specialrule{.05em}{0em}{.2em} 
		\makecell{SR \\ ER \\ SC \\ EC \\ CC \\ CL} & \makecell{5 \\ 0.05 \\ 0.5 \\ 0.05 \\ 0.5 \\ 0.5} \makecell{$10^3$ \\ $10^4$ \\ $10^3$ \\ $10^4$ \\ $10^4$ \\ $10^4$} &    \makecell{3 \\ 0.05 \\ 0.1 \\ 0.1 \\ 0.01 \\ 0.01} \makecell{$10^3$ \\ $10^3$ \\ $10^3$ \\ $10^3$ \\ $10^3$ \\ $10^3$} &  \makecell{1 \\ 2 \\ 1 \\ 1 \\ 0.5 \\ 0.5} \thickspace\makecell{0.1 \\ 1 \\ 0.1 \\ 0.1 \\ 3 \\ 3}\thickspace\makecell{500 \\ $10^3$ \\ $10^3$ \\ $10^3$ \\ $5\times10^3$ \\ $5\times10^3$}\thickspace\makecell{5 \\ 0.1 \\ 0.1 \\ 0.1 \\ 1 \\ 1}\\ [.2em] 
		\specialrule{.05em}{0em}{.2em}
	\end{tabular}
\end{table}
%%%%%%%%%%%%%%%%%%%%%%%%%%%%%%%%%%%%%%%%%%%%%%%%%%%%%%%%%%%%%%%%%%%%%%%%%%%%%%%%%%%%%%%%%%%%%%%%%%%%%%%
\bibliographystyle{IEEEbib}
\bibliography{refs}
\end{document}